\documentclass[a4paper]{article}

\usepackage[utf8]{inputenc} % set input encoding (not needed with XeLaTeX)

\usepackage{amsfonts}
\usepackage{authblk}
\usepackage{amssymb}
\usepackage{amsbsy}
\usepackage[margin=2cm]{geometry}

\usepackage{xcolor}
\usepackage{cite}
\usepackage{mathrsfs}
\usepackage{hyperref}
\usepackage{graphicx} % support the \includegraphics command and options

\usepackage{booktabs} % for much better looking tables
\usepackage{array} % for better arrays (eg matrices) in maths
\usepackage{paralist} % very flexible & customisable lists (eg. enumerate/itemize, etc.)
\usepackage{verbatim} % adds environment for commenting out blocks of text & for better verbatim
\usepackage{subfigure}
\makeatletter
\renewcommand{\@thesubfigure}{\normalsize(\textbf{\alph{subfigure}})}
\makeatother

\usepackage{fancyhdr} % This should be set AFTER setting up the page geometry
\title{Qubit representation of qudit states: correlations and state reconstruction.}
\author[1,2]{Julio A. L\'opez-Sald\'ivar}
\author[1]{Octavio Casta\~nos}
\author[3]{Margarita A. Man'ko}
\author[2,3,4]{Vladimir I. Man'ko}
\affil[1]{Instituto de Ciencias Nucleares, Universidad Nacional Aut\'onoma de M\'exico,
Apdo. Postal 70-543, Ciudad de México 04510, M\'exico}
\affil[2]{Moscow Institute of Physics and Technology (State University), Institutskii per. 9,
Dolgoprudnyi, Moscow~Region 141700, Russia}
\affil[3]{Lebedev Physical Institute, Russian Academy of Sciences, Leninskii Prospect 53,
Moscow 119991, Russia}
\affil[4]{Department of Physics, Tomsk State University, Lenin Avenue 36, Tomsk 634050, Russia}
\date{}

\begin{document}
%\linenumbers




\maketitle

\begin{abstract}
A method to establish a qubit decomposition of a general qudit state is presented. This new representation allows a geometrical depiction of any qudit state in the Bloch sphere. Additionally, we show that the nonnegativity conditions of the qudit state imply the existence of quantum correlations between the qubits which compose it.  These correlations are used to define new inequalities which the density matrices components must satisfy. The importance of such inequalities in the reconstruction of a qudit state is addressed. As an example of the general procedure the qubit decomposition of a qutrit system is shown, which allows a classification of the qutrit states by fixing their invariants ${\rm Tr}(\hat{\rho}^2)$, ${\rm Tr}(\hat{\rho}^3)$. 
%\keywords{qudit states \and geometrical representation of quantum states \and quantum correlations \and quantum state reconstruction \and Bloch sphere}
\end{abstract}

\section{Introduction}
Recently, it has been shown that the qutrit state density matrix can define different qubit density matrices \cite{chernega17a,chernega17b,chernega17c}. This is done by making an extension of the qutrit system into a two-qubit state and then performing the partial trace to obtain qubit states. The resulting qubits can be represented geometrically by the definition of triangles and a set of squares which can be expressed in terms of classical probability distributions. Additionally, the sum of the areas of these figures can be used to determine the quantum entanglement in the two-qubit case \cite{entropy18}. 

It is known that a general qubit state has a geometric representation as a point inside the Bloch sphere. In this representation the distance from the center of the sphere and the point representing the qubit is the purity of the system. Several ways to extend this geometrical representation to a qutrit or qudit state have been proposed \cite{ellipsoid,geo-qutrit,kimura,mendas,kowski,majorana}.

Also, the state reconstruction of $s$-dimensional spin systems is regularly done using tomographic measurements which lead to the calculation of the density matrix components. The number of measurements needed is $4 s+1$ \cite{weigertbook}. However, it has been shown that an estimate of the density matrices can be obtained without all the information \cite{buzek}.

The importance of the characterization of qudit systems has been of interest in recent years. They have been used in the construction of a family of Bell type inequalities \cite{collins}, to define large alphabet key distributions \cite{ali}, to enhance the sensitivity of photodetectors \cite{lloyd}, to emulate spin systems \cite{neeley}, in telecloning processes \cite{araneda}, and coherence characterization \cite{maziero}. It has been established that a single qudit quantum algorithm is enough to have more efficiency than any classical counterpart \cite{gedik}. A way to generate entangled qudit states on a chip has been found in \cite{kues}. A universal qudit quantum computation using only three wave mixing interactions has been reported in \cite{niu}. It has been shown that the quantum logic used for qubits can be simplified using qudits \cite{lanyon}. Also, a minimal set of measurements in qudit state tomography has been given in \cite{ha}.

The main goal of this work is to establish a general procedure to define an ensemble of qubit density matrices for the description of any qudit system. We also use this ensemble of qubits to establish bounds to the probabilities and coherences for the qudit system, specially for the reconstruction of a qutrit system.  

Our work is organized as follows: In section 2, the qutrit is decomposed in terms of three different sets of qubits. In order to show the correlation between the different resulting qubits we calculate the correlation between eigenvalues for two different qubits. A new geometric representation of an arbitrary qutrit system in terms of three points on the Bloch sphere is proposed in section 3. Also in this section, a classification of the zones of this geometrical representation is given in terms of the invariants ${\rm Tr (\hat{\rho}^2)}$, ${\rm Tr (\hat{\rho}^3)}$ of the qutrit system. 
In section 4, a generalization of the procedure to describe any qudit system in terms of an ensemble of qubits is presented, including its geometric representation in terms of $d$ points in the Bloch sphere. In section 5, the positivity conditions for the ensemble of qubits describing a qutrit state are used to establish bounds in 
the unknown components of its density matrix, that is, when in an experiment only partial information is provided for the state. The extension of the procedure for a qudit system is also discussed. Finally, some conclusion are given.

\section{Qubit decomposition of a qutrit state }
In this section, we establish the procedure to calculate the qubit decomposition of a general qutrit system. We call this procedure qubit decomposition as at the end we can rewrite the qutrit system in terms of qubit parameters. To show that the resulting qubits are correlated, as they came from the same qutrit system, we study the Pearson correlation between two of them.

A general qutrit state can be defined in terms of the Bloch vector parametrization \cite{gellmann} as follows:
\begin{equation}
\hat{\rho}=\frac{1}{2}\left( \begin{array}{ccc}
\frac{2}{3}+a_7+\frac{a_8}{\sqrt{3}} & a_1-i a_4 & a_2-i a_5 \\
a_1+i a_4 &  \frac{2}{3}-a_7+\frac{a_8}{\sqrt{3}} & a_3 - i a_6 \\
a_2+i a_5 & a_3 +i a_6 & \frac{2}{3}-2\frac{a_8}{\sqrt{3}}
\end{array}\right) \, ,
\label{qutrit}
\end{equation}
where the vector $\mathbf{a}=(a_1, \cdots,a_8)$ is called the generalized Bloch vector of the system. To obtain the qubit decomposition, one first map the $3 \times 3$ density matrix onto different $4\times 4$ matrices with the same eigenvalues as the original state plus one eigenvalue equal to zero, this guarantees the nonnegativity of the extended density matrices as $\hat{\rho}$ has nonnegative eigenvalues and we only add an eigenvalue equal to zero. This lead us to the definition of states with one row and one column equal to zero as follows:
\begin{eqnarray}
\hat{\sigma}^{(1)}&=& \left( \begin{array}{cc}
\hat{\rho}_{3\times3} & 0_{3\times 1} \\
0_{1\times 3} & 0
\end{array}\right) \, , \
\hat{\sigma}^{(2)}= \left( \begin{array}{cc}
0_{1\times 3} & 0 \\
0_{3\times 1} & \hat{\rho}_{3\times3}
\end{array}\right) \, , \nonumber \\
\hat{\sigma}^{(3)}&=&\left(\begin{array}{cccc}
\rho_{11} & 0 & \rho_{12} & \rho_{13} \\
0 & 0 & 0 & 0 \\
\rho_{21} & 0 & \rho_{22} & \rho_{23} \\
\rho_{31} & 0 & \rho_{32} & \rho_{33}
\end{array}\right) , \
\hat{\sigma}^{(4)}=\left(\begin{array}{cccc}
\rho_{11}  & \rho_{12}& 0 & \rho_{13} \\
\rho_{21}  & \rho_{22}& 0 & \rho_{23} \\
0 & 0 & 0 & 0 \\
\rho_{31}  & \rho_{32}& 0 & \rho_{33}
\end{array}\right) \, .
\label{mat}
\end{eqnarray}
Once these four density matrices are constructed, each one of them should be seen as a composed density matrix by a two-qubit system written on the basis $\vert m_1=\pm 1/2, m_2=\pm 1/2 \rangle$. Then, using the partial trace, six different qubit density matrices ($\hat{\rho}_1, \cdots, \hat{\rho}_6$) can be obtained , i.e.,
\begin{eqnarray}
\hat{\rho}_1=\left( \begin{array}{cc}
1-\rho_{33} & \rho_{13} \\
\rho_{31} & \rho_{33}
\end{array}\right) \, , \quad 
\hat{\rho}_2=\left( \begin{array}{cc}
1-\rho_{22} & \rho_{12} \\
\rho_{21} & \rho_{22}
\end{array}\right) \, , \nonumber \\ 
\hat{\rho}_3=\left( \begin{array}{cc}
\rho_{11} & \rho_{13} \\
\rho_{31} & 1-\rho_{11}
\end{array}\right) \, , \quad 
\hat{\rho}_4=\left( \begin{array}{cc}
\rho_{22} & \rho_{23} \\
\rho_{32} & 1-\rho_{22}
\end{array}\right) \, , \nonumber \\
\hat{\rho}_5=\left( \begin{array}{cc}
\rho_{11} & \rho_{12} \\
\rho_{21} & 1-\rho_{11}
\end{array}\right) \, , \quad 
\hat{\rho}_6=\left( \begin{array}{cc}
1-\rho_{33} & \rho_{23} \\
\rho_{32} & \rho_{33}
\end{array}\right) \, .
\label{qubits1}
\end{eqnarray}
From the positivity conditions of $\hat{\rho}$, one can see that the previous matrices are nonnegative, hermitian and with trace equal to one. To prove this, we note that the matrices $\hat{\sigma}^{(1)},\ldots,\hat{\sigma}^{(4)}$, satisfy the conditions to be  bona fide density matrices, e.g., these density matrices can describe a 4 level system where the transitions to one of the levels are forbidden, due to energetic issues or because there exist selection rules which forbid its occupation. For each one of the density matrices, the partial trace operation gives two qubit density matrices, then $\hat{\rho}_1,\ldots,\hat{\rho}_6$ are valid density matrices. Of course, the previous partial density matrices must fulfill the standard nonnegativity conditions for a qubit state: $1/2\leq{\rm Tr}(\hat{\rho}^2_j)\leq1$, or equivalently $0\leq \det \hat{\rho}_j \leq 1/4$. Also, the fidelity~\cite{jozsa} given by
\begin{equation}
F(\hat{\rho}_j, \hat{\rho}_k) = {\rm Tr}(\hat{\rho_j} \hat{\rho_k})+2\sqrt{\det{\hat{\rho}_j} \det{\hat{\rho}_k}} \, ,
\end{equation}
provides other conditions over the qutrit state components, $0 \leq F(\hat{\rho}_j, \hat{\rho}_k)\leq 1$ ($j,k=1, \ldots , 6$, $j \neq k$). Furthermore, the qubit matrices must also satisfy any other conditions for a $2 \times 2$ state as the ones provided by the Renyi \cite{renyi}, Tsallis \cite{tsallis}, or von Neumann \cite{neumann} entropies, together with the conditions given by the mutual information, as the mentioned qubits are obtained from the reduction of a $4\times 4$ system. All these inequalities can be used to establish bounds for the different density matrix components in the case of a state reconstruction experiment where not all the measurements have been performed.

\begin{figure}
\centering
\includegraphics[scale=0.25]{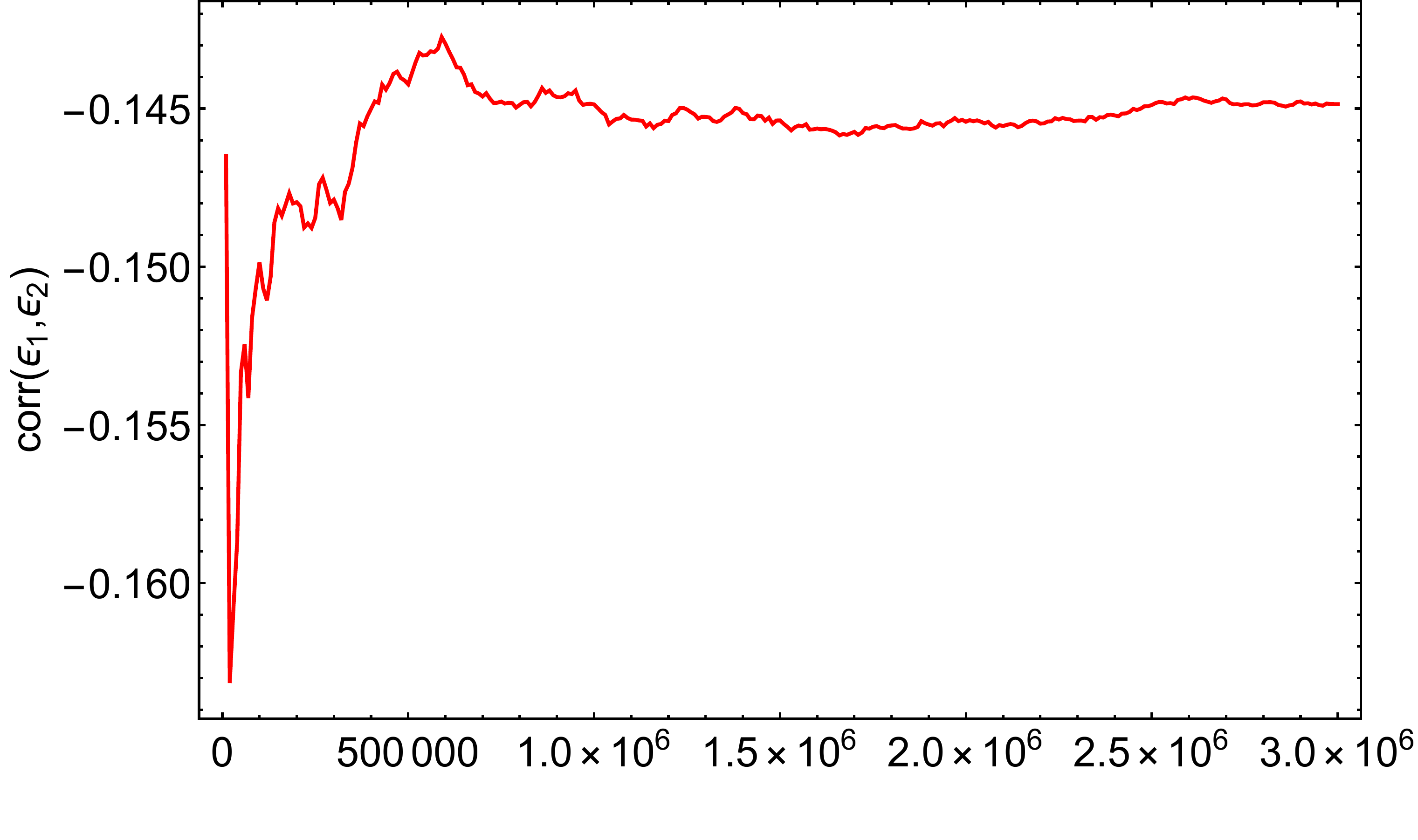}
\includegraphics[scale=0.25]{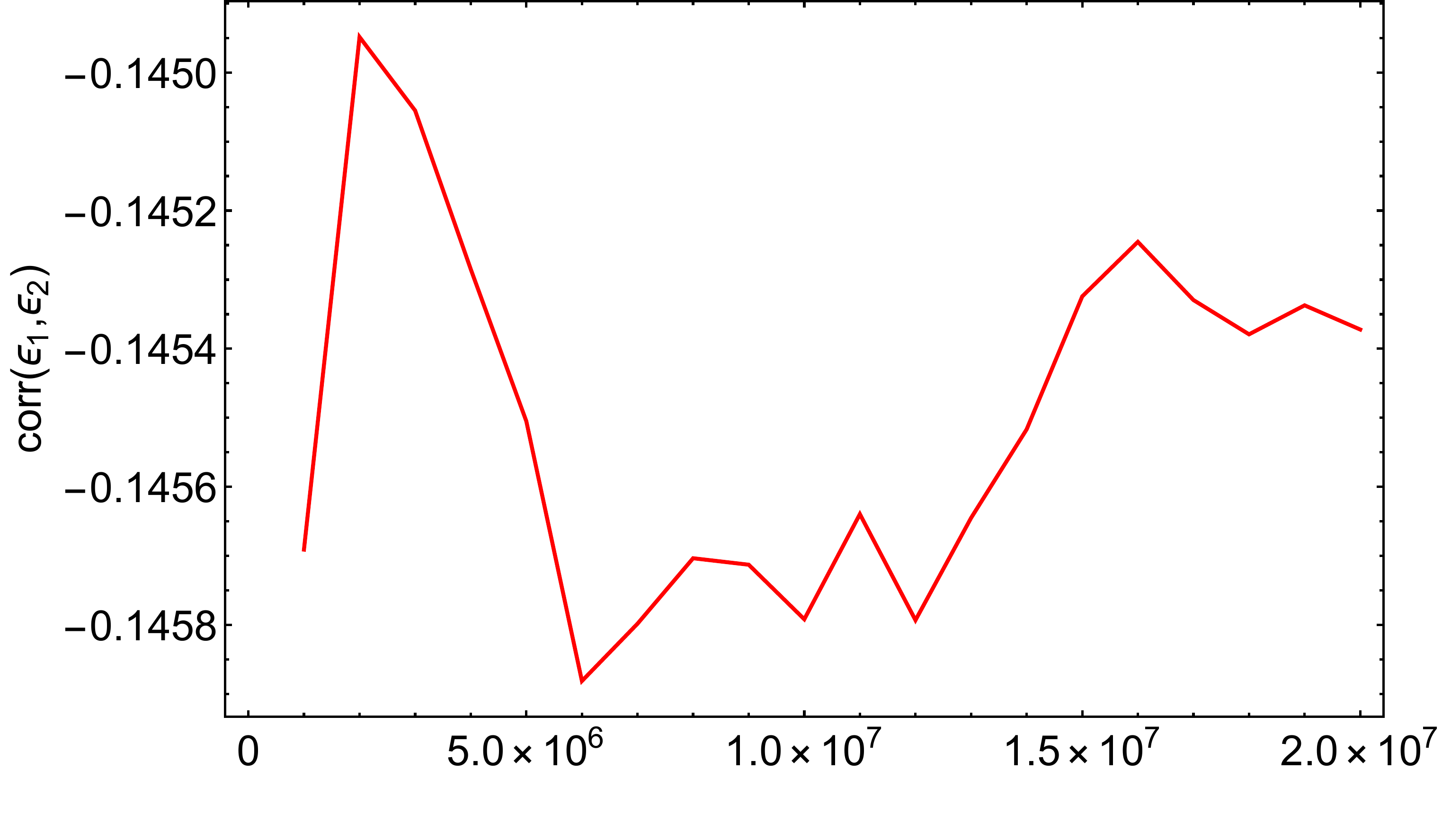}
\caption{Pearson correlation for the smallest eigenvalues of the qubit systems $\hat{\rho}_1$ ($\epsilon_1$) and $\hat{\rho}_2$ ($\epsilon_2$). The horizontal axis is the number of random generated qutrit states $\hat{\rho}$ taken into account to calculate the correlation. \label{corr}}
\end{figure}

As the qubit matrices are defined by means of the entries of the initial qutrit state $\hat{\rho}$, then there exist correlations between them. For example, we can think on a operation (unitary or not) which transforms the state $\hat{\rho}$ onto the state $\hat{\rho}'$. Suppose an hypothetical, nonunitary transformation which mainly changes one of the off-diagonal components of the density matrix, in other words only some of the qubits defined above are affected a priori. However, as the conditions of nonnegativity and trace of the density matrix must be still valid then one can see that all the qubits must be changed in such a form to preserve the validity of all the conditions; so a change in one affects all others. These correlations can be measured using the standard classical definition, e.g., the correlations between one of the eigenvalues of the qubit density matrices $\hat{\rho}_1$ ($\epsilon_1$), and $\hat{\rho}_2$ ($\epsilon_2$) can be obtained through the Pearson correlation~\cite{pearson}
\begin{equation}
{\rm corr}(\epsilon_1, \epsilon_2)=\frac{\langle \epsilon_1 \epsilon_2 \rangle-\langle \epsilon_1 \rangle \langle \epsilon_2 \rangle}{\sigma_{\epsilon_1} \sigma_{\epsilon_2}} \, ,
\label{corre}
\end{equation}
where $\sigma_{\epsilon_1}$ and $\sigma_{\epsilon_2}$ are the standard deviations of the eigenvalues $$\sigma_\zeta=\sqrt{\langle \zeta^2 \rangle-\langle \zeta \rangle^2}\, .$$ The mean value $\langle \cdots \rangle$ is taken over all the possible values of the qubits $\hat{\rho}_1$, $\hat{\rho}_2$ defined from a random choice of the qutrit state $\hat{\rho}$. Although, the qutrit state is taken randomly from all the possible states in the Hilbert space, it is possible to see that the correlation tends to a certain number. Here, we notice again that the two qubits $\hat{\rho}_1$ and $\hat{\rho}_2$ are defined by the entries of the same qutrit density matrix $\hat{\rho}$. Then one can consider them as two interacting qubits where the nonnegativity conditions of the qutrit play the role of the interaction between them. Because if $\hat{\rho}$ varies then $\hat{\rho}_1$ and $\hat{\rho}_2$ must vary in order to satisfy the nonnegativity conditions. The correlation between the smallest eigenvalues of both qubits should be determined by these interactions, if one choose the state $\hat{\rho}$ randomly one can think that the correlation might result in a random correlation but instead an almost convergent distribution is obtained. In fig.~(\ref{corr}) one can see the tendency of the correlation~(\ref{corre}) between the smallest eigenvalues of  $\hat{\rho}_1$ ($\epsilon_1$) and $\hat{\rho}_2$ ($\epsilon_2$) as a function of the number of different states $\hat{\rho}$ which are generated randomly. Here we show that the value of the correlation tends to a value $\approx -0.145$ and has certain small fluctuations around that number due to the quantum nature of the system. This value means anticorrelation between the mentioned eigenvalues, i.e., as one increase the value of $\epsilon_1$ then the value of $\epsilon_2$ decrease. In the same figure it can be seen that despite the large number of random states generated $\approx 20$ millions, a certain variation in the correlation can be noted, we think this behavior is due to the fact that we are measuring the correlation between two quantum objects. We point out that the same behavior is present for any other correlation of the different eigenvalues of the density matrices $\hat{\rho}_1, \ldots , \hat{\rho}_6$.

\section{Geometric representation of a qutrit system.}
Here, we propose a new representation of any qutrit state by means of three points in the Bloch sphere. The three points are obtained using the Bloch representation of three different qubit density matrices of the six possible as stated in Eq.~(\ref{qubits1}). Some of the qutrit properties as the purity and the coherence of the state are then studied, and a geometrical interpretation in the Bloch sphere for these quantities is proposed. Finally, a classification of the zones of the geometrical representation of the qutrit in terms of the state invariants ${\rm Tr}(\hat{\rho}^2)$ and ${\rm Tr}(\hat{\rho}^3)$ is discussed.
%Figure 1

\begin{figure}
\centering
\subfigure[]
{\includegraphics[scale=0.25]{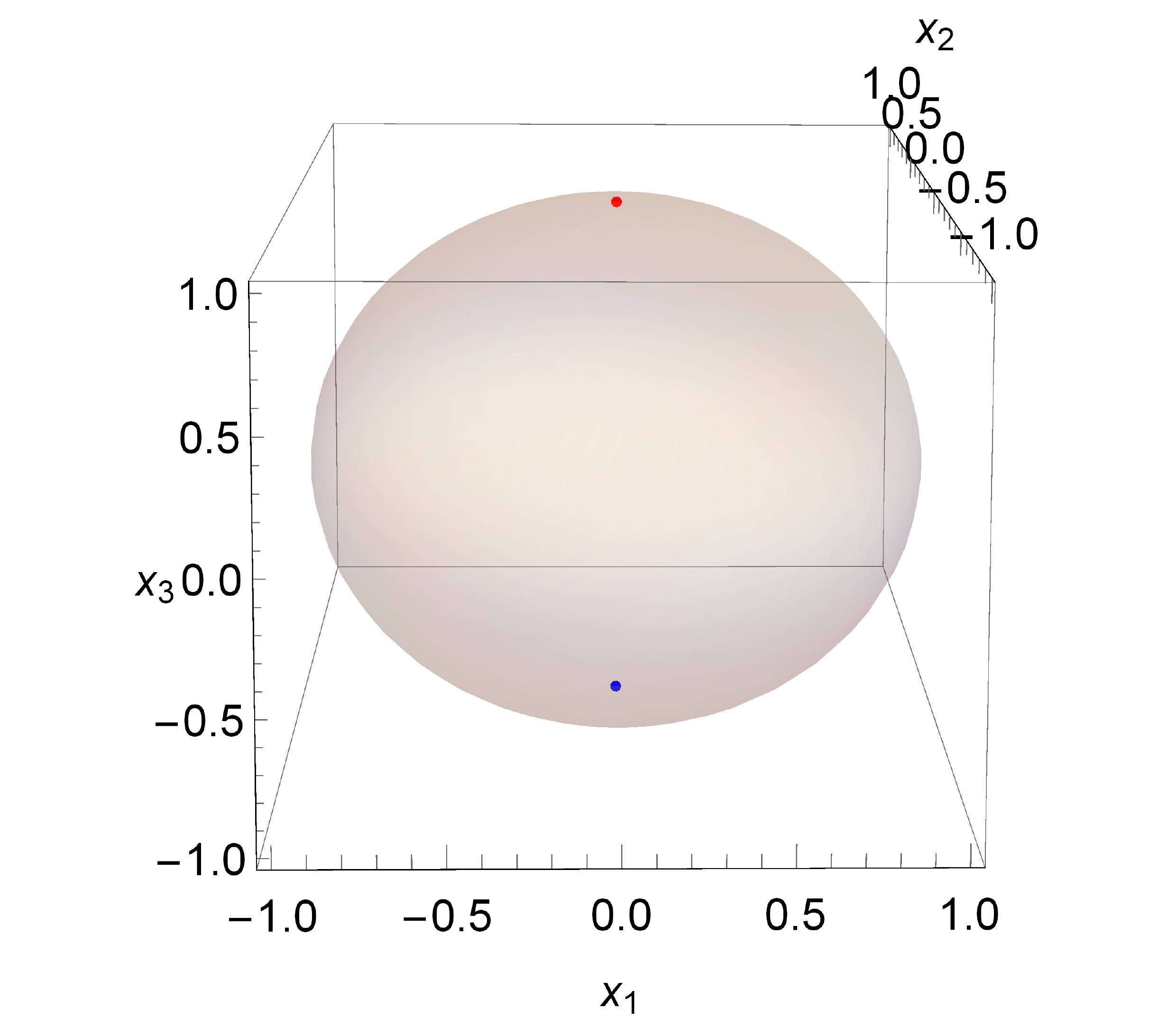} \label{fig1a}}
\subfigure[]
{\includegraphics[scale=0.24]{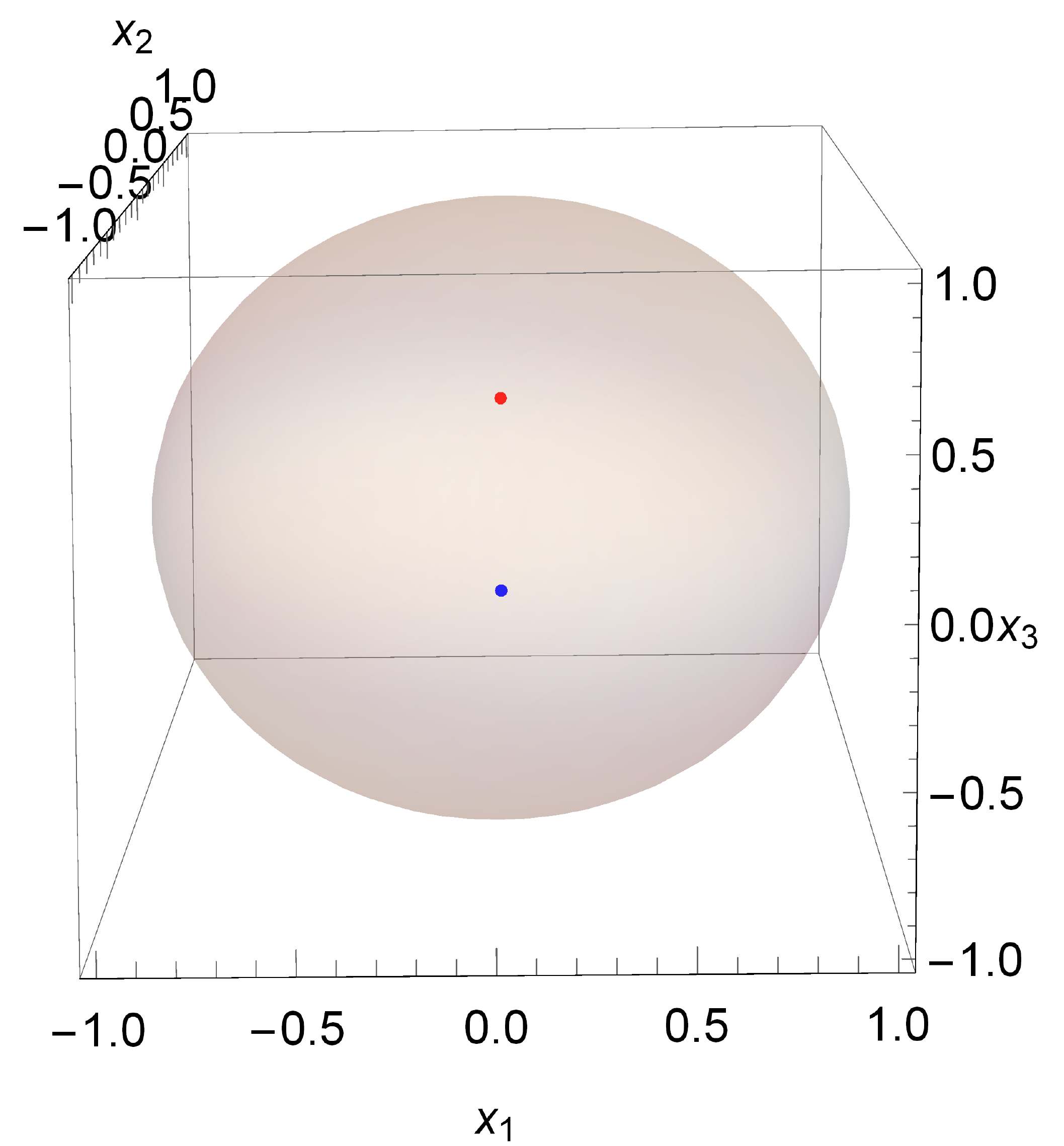} \label{fig1b}}
\subfigure[]
{\includegraphics[scale=0.24]{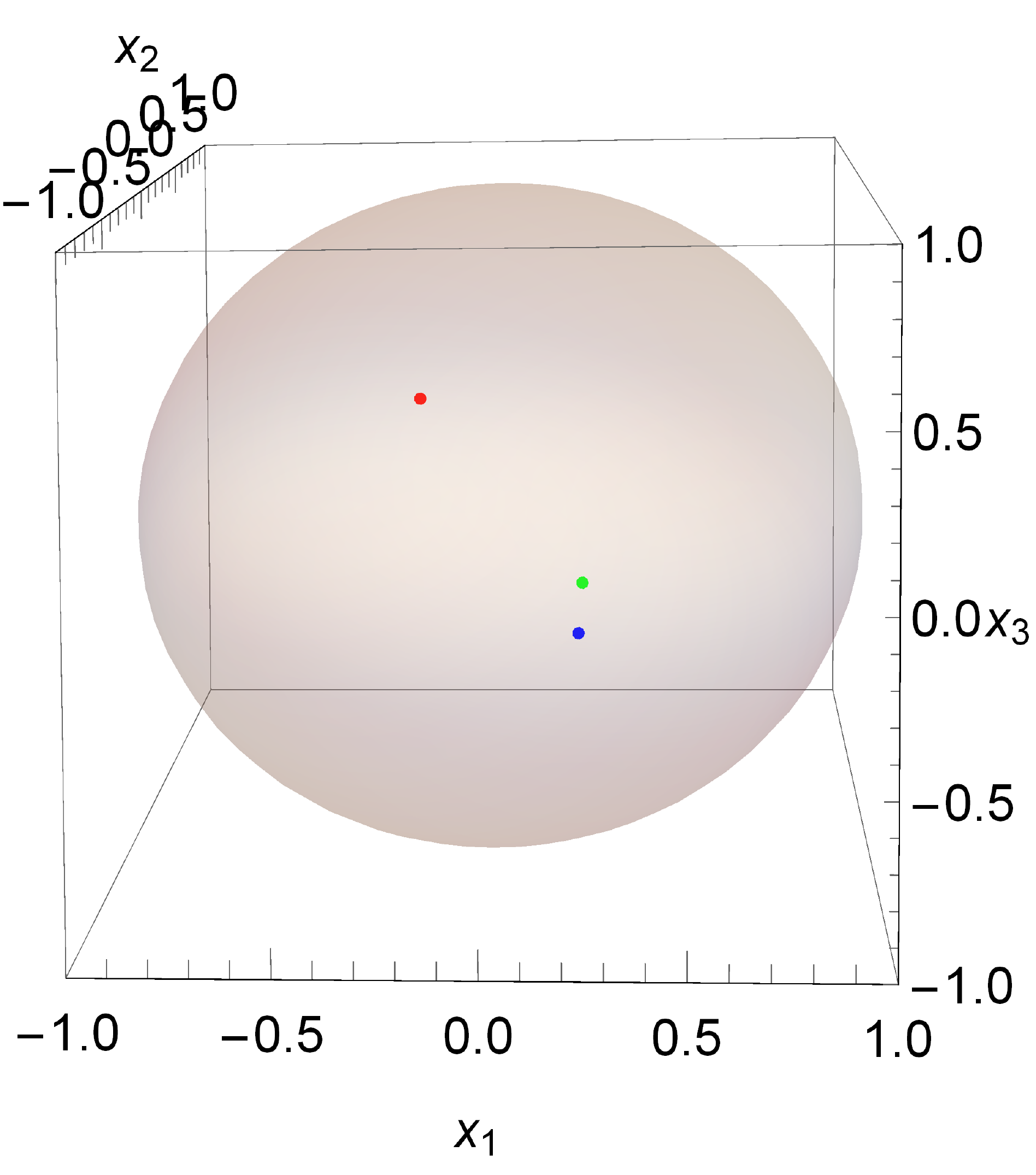} \label{fig1c}}
\caption{Graphical representation of (a) the pure state $\vert j=1,m=1 \rangle$, (b) the most mixed state $\hat{\rho}=\hat{\mathbf{I}}/3$, and (c) for the particular state defined in Eq.~(\ref{part}).}
\end{figure}

Given that the qutrit state has 8 independent variables, one can take a set of 3 qubits which must contain all the off-diagonal terms of the original state $\hat{\rho}$. Then using the correspondence between the Bloch vectors of the set and the generalized Bloch vector of the qutrit, one can define a graphical representation of the system. This geometrical representation of the qutrit is given by the Bloch sphere representation of its 3 qubit decomposition. Consider a quorum of qubit density matrices which allow us to describe a qutrit system, e.g., choosing the sets $\{ \hat{\rho}_1, \hat{\rho}_2, \hat{\rho}_4\}$, $\{\hat{\rho}_3,\hat{\rho}_5,\hat{\rho}_6\}$, or $\{\hat{\rho}_1, \hat{\rho}_4, \hat{\rho}_5 \}$, etc. The graphical representation of the Bloch vector for every state of the quorum is determined by a point in the Bloch sphere. For the all the density operators $\hat{\rho}_{j_k}$ which form the quorum $\{\hat{\rho}_{j_1}, \hat{\rho}_{j_2}, \hat{\rho}_{j_3} \}$, we will have the three dimensional point ${\cal P}_k=((\hat{\rho}_{j_k})_{12}+(\hat{\rho}_{j_k})_{21},i((\hat{\rho}_{j_k})_{12}-(\hat{\rho}_{j_k})_{21}),2(\hat{\rho}_{j_k})_{11}-1)$,  ($k=1,2,3$). It is important to address that the resulting 3 points must be differentiated (e. g. must have RGB coloring or different numbers)  in order to distinguish different qutrit states represented  with the same points in different order. The different sets of three points give us information of the qutrit system. In fig.~\ref{fig1a} one can see the graphical representation of the pure state $\vert j=1,m=1 \rangle$, in fig.~\ref{fig1b} the most mixed state $\hat{\rho}=\hat{\mathbf{I}}/3$ is represented. In both cases, we have only two different points $\mathcal{P}_1$, $\mathcal{P}_2$ of the three possible, these cases are not typical, as in most of the states, we can define three different points as it is shown in fig.~(\ref{fig1c}) for the state 
\begin{equation}
\hat{\rho}=\frac{1}{9}\left( \begin{array}{ccc}
3 & 1+i & -1 \\
1-i & 3 & 1-i \\
-1 & 1+i & 3
\end{array}
\right) \, .
\label{part}
\end{equation}
The representation presented in figs. \ref{fig1a}-\ref{fig1c} was made using the Bloch vectors of the qubit density matrices $\{\hat{\rho}_1,\hat{\rho}_4,\hat{\rho}_5\}$ of Eq.~(\ref{qubits1}), which we propose as the canonical way to obtain the Bloch representation of the qutrit. Because for this canonical set the purity and the coherence of the system have a geometrical interpretation, and this feature is not present in the other possible representations. In this representation, one can demonstrate that the sum of the qubit purities is always larger than the purity of the qutrit state, i.e.,
\[
{\rm Tr}(\hat{\rho}^2)= \sum_{j=1,4,5} \left({\rm Tr}(\hat{\rho}_j^2) -\frac{x_3^{(j)2}}{4}\right)-\frac{5}{4}\, ,
\]
where $x_3^{(j)}$ is the third component of the Bloch vector associated to $\hat{\rho}_j$. In particular one can see that ${\rm Tr}(\hat{\rho}^2)\leq \sum_{j=1,4,5} {\rm Tr}(\hat{\rho}_j^2)$, so an upper bound for the purity of the system can be given by the sum of the distances from the center of the Bloch sphere to each one of the three points given by the Bloch vectors of $\hat{\rho}_1$, $\hat{\rho}_4$, and $\hat{\rho}_5$ minus the sum of the squared vertical components. This property lead us to have a way to compare two different states and guess which one has a larger purity. This can be checked for the states of fig.~\ref{fig1a}- \ref{fig1c}. It is also noteworthy that the coherence of the system ($C(\hat{\rho})=\sum_{j,k(j\neq k)} \vert \rho_{jk} \vert$) can be expressed as the sum of the coherences of the qubits

%Figure 2

\[
C(\hat{\rho})= \sum_{j=1,4,5} C(\hat{\rho}_j) \, .
\]
This coherence can be obtained from the geometrical representation by summing the distances of the three points $\mathcal{P}_k$, $k=1,2,3$ to the vertical axis $x_3$. From this, one can conclude that all diagonal states will always be represented by collinear points over the vertical axis.

\begin{figure}
\centering
\subfigure[]
{\includegraphics[scale=0.27]{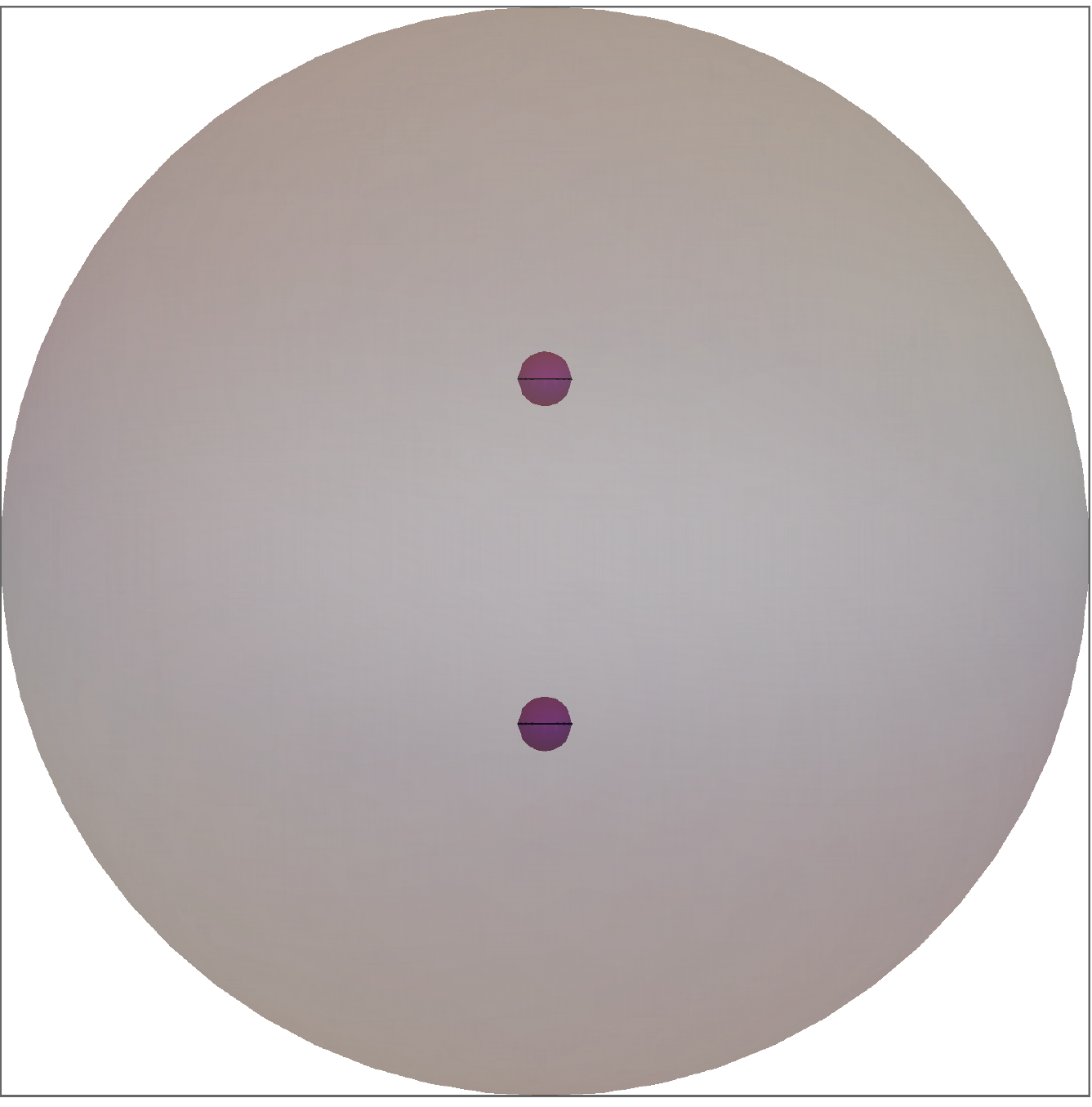} \label{fig3a}} \
\subfigure[]
{\includegraphics[scale=0.27]{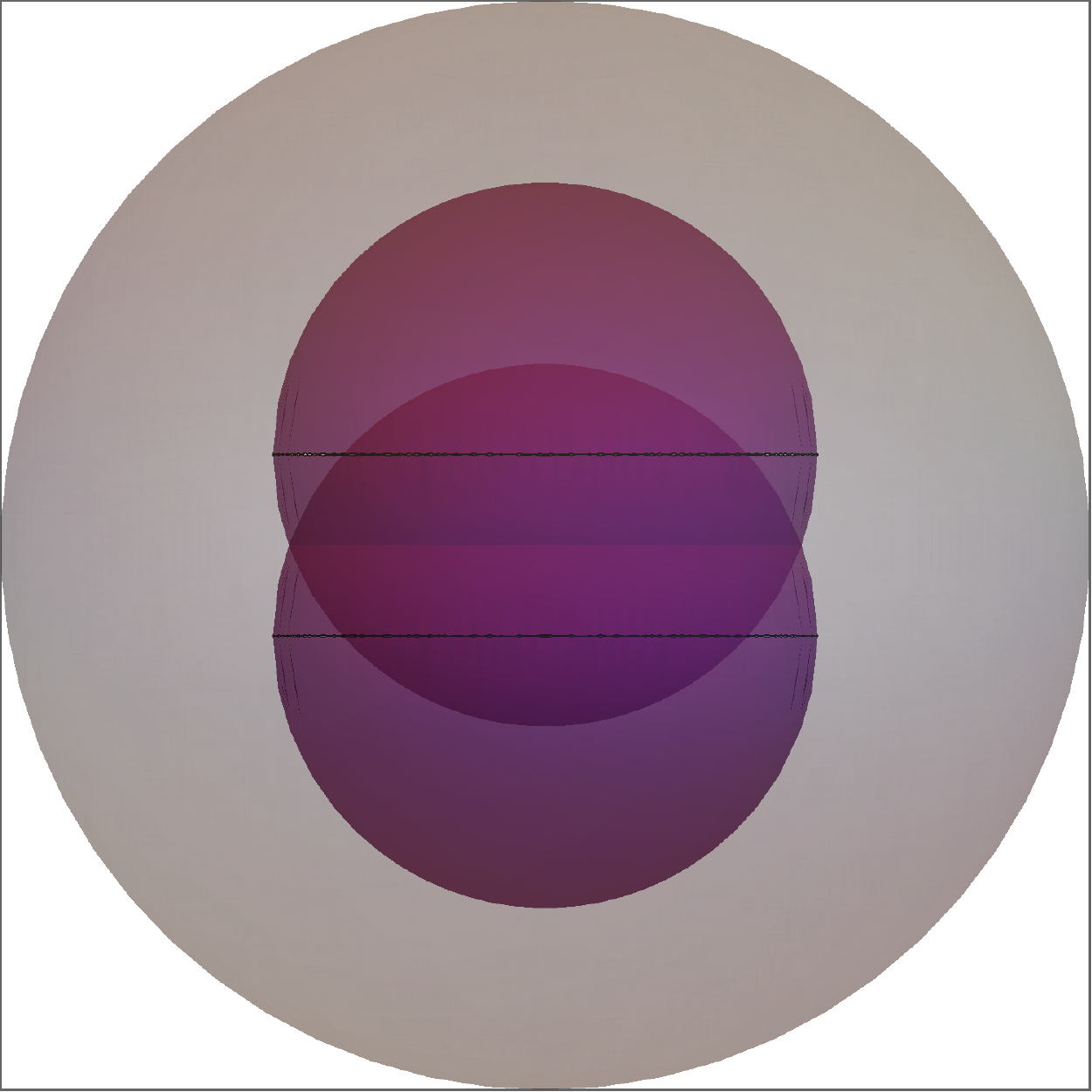} \label{fig3b}} \
\subfigure[]
{\includegraphics[scale=0.27]{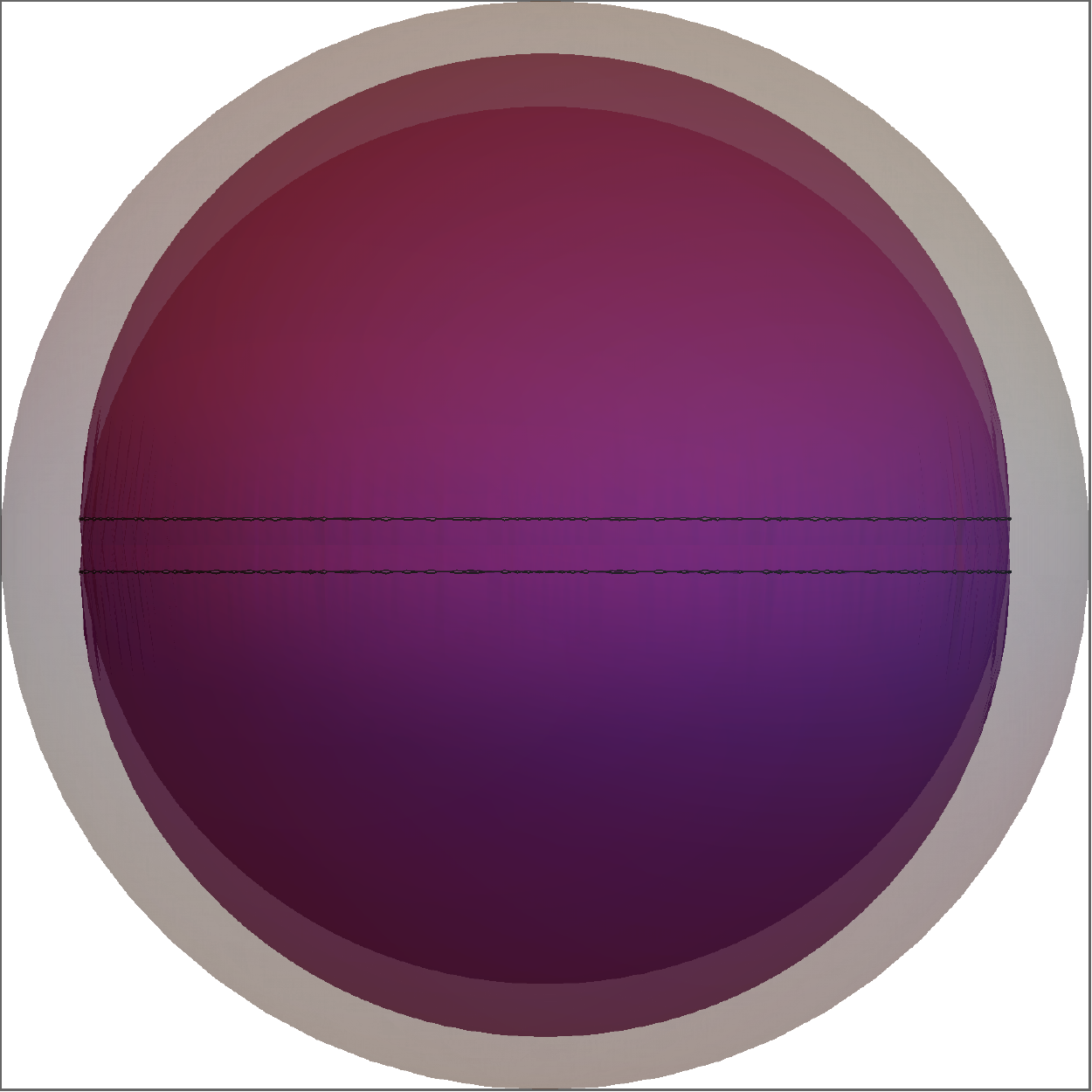} \label{fig3c}}
\caption{Allowed zones for the points $\mathcal{P}_1$, $\mathcal{P}_2$, and $\mathcal{P}_3$ for (a) a qutrit state with purity ${\rm Tr}(\hat{\rho}^2)=67/200$, (b) purity ${\rm Tr}(\hat{\rho}^2)=1/2$, and (c) ${\rm Tr}(\hat{\rho}^2)=41/50$. In all these cases the regions are evaluated at the maximal allowed value of $t_3=\frac{1}{18} \left(-\sqrt{6 t_2-2}+3 t_2 \left(\sqrt{6 t_2-2}+6\right)-4\right)$}
\end{figure}

The three points depicting the state $\hat{\rho}$ can only be located in a certain region of the Bloch sphere once the invariants of the state (${\rm Tr}(\hat{\rho}^2)=t_2$, ${\rm Tr}(\hat{\rho}^3)=t_3$) are fixed, this property can be used to do a classification of the general qutrit state. If the purity of the system is equal to ${\rm Tr}(\hat{\rho}^2)=1/3$ (the most mixed state), then the three points corresponding to the Bloch vectors of $\hat{\rho}_1$, $\hat{\rho}_4$, $\hat{\rho}_5$ can only be $\mathcal{P}_1=(0,0,1/3)$, $\mathcal{P}_2=\mathcal{P}_3=(0,0,-1/3)$. When we have a state with larger purity but close to $1/3$ the points representing the states can only lie in two spherical regions with center around the points denoting the most mixed state $(0,0,\pm1/3)$. As we increase the purity, the radius of the spherical regions becomes larger until the Bloch sphere is completely full at a purity equal to 1. This behavior can be seen in figs.~\ref{fig3a}-\ref{fig3c} . In general, given the invariants ${\rm Tr}(\hat{\rho}^2)=t_2$, ${\rm Tr}(\hat{\rho}^3)=t_3$ the regions where the points $\mathcal{P}_k$, $k=1,2,3$ lie can be obtained solving the system of inequalities in terms of the qubit density matrices
\begin{eqnarray}
1/2\leq {\rm Tr}(\hat{\rho}_1) \leq 1, \quad 1/2\leq {\rm Tr}(\hat{\rho}_4) \leq 1, \nonumber \\
1/2\leq {\rm Tr}(\hat{\rho}_5) \leq 1, \quad \det \hat{\rho}\geq 0 \label{t2t3}
\end{eqnarray}
together with the equations that fix the invariants $t_2$, $t_3$ to constant values. These expressions lead to the allowed zones, two spherical regions symmetric with respect to the $x-y$ plane. The spheres are centered at $(x_0=0, y_0=0, \pm z_0)$, where $z_0$ is restricted to be in a plane whose borders are the curves 
\[
z_0=\left\{ \frac{1}{3} + \sqrt{\frac{t_2}{6}-\frac{1}{18}} \, ,  \frac{1}{3} - \sqrt{\frac{t_2}{6}-\frac{1}{18}} \, , \frac{1}{2}-\sqrt{\frac{t_2}{2}-\frac{1}{4}} \right\}\, .
\]
These limiting curves can be obtained by considering the minimum $t_{3 {\rm min}}$ and maximum $t_{3 {\rm max}}$ into the inequalities~(\ref{t2t3}). These values are given by \cite{kimura}
\begin{eqnarray}
t_{3 {\rm max}}&=&\frac{1}{18} \left\{-\sqrt{6 \, t_2-2} + 3 \, t_2 \left(\sqrt{6 \, t_2-2}+6\right)-4\right\} \, , \nonumber \\
t_{3 {\rm min}} &=& \Bigg\{ \begin{array}{cc} \frac{1}{18} \left(\sqrt{6\, t_2-2} - 3 \,t_2 \left(\sqrt{6 \,t_2-2}-6\right)-4\right)\quad {\rm for}  \quad \frac{1}{3}\leq t_2\leq \frac{1}{2} \, ,  \\
\frac{1}{2} (3 \, t_2-1) \quad {\rm for} \quad  \frac{1}{2}\leq t_2\leq 1 \, .
\end{array}
\end{eqnarray}
The spheres have the same radius with values bounded by the following curves,
\[
R=\Bigg\{\sqrt{\frac{3 \, t_2}{2}-\frac{1}{2}},\frac{1}{2}+ \sqrt{\frac{t_2}{2}-\frac{1}{4}},\sqrt{2 \, t_2-\frac{2}{3}} \Bigg\} \, ,
\]
where the first two curves are obtained from the solution of Eq.~(\ref{t2t3}) with the invariant $t_3$ taken as $t_{3max}$, and the second part of $t_{3min}$ respectively. The third curve can be obtained replacing the equation which fixes $t_3$ by the inequality $t_{3max}\leq t_3 \leq t_{3min}$.

The qubit decomposition and the graphical representation described above can be generalized for any qudit system, this is discussed in the following section.

%Figure 3

\section{Qubit decomposition for higher dimensions.}

In this section the procedure of qubit decomposition is extended to a general qudit system. To obtain this decomposition the idea of the extension of the density matrix to higher dimensions is used. As in the qutrit case we also use the partial trace of the extended density matrices as the partial trace operation always give us a group of density operators. The generalization of the geometrical representation of the qutrit state studied above is also addressed.

The general procedure to obtain the qubit decomposition of a qudit state is the following:
\begin{enumerate}
\item
The density matrix $\hat{\rho}$ of the $d$-dimensional qudit system must be extended to all the possible density matrices $\hat{\sigma}^{(r)}$ of dimension $l>d$, where $l$ is the closest even number larger than $d$ (i.e., $l=d+2$ for even $d$, or $l=d+1$ for odd $d$). The resulting matrices must contain all the elements of $\hat{\rho}$ and up to two rows and two columns equal to zero (one row and one column for odd $d$ and two rows and two columns for even $d$). The zero rows and columns must be taken in pairs of row and column with the same position in the list of numbers $1,\cdots,l$, in order to assure that ${\rm Tr}\,\hat{\sigma}^{(r)}=1$. E.g., for $d=4$, $\hat{\sigma}^{(r)}$ are the matrices of dimension $6$ which contains 2 rows and 2 columns equal to zero. For $d=5$, $\hat{\sigma}^{(r)}$ must be also $6\times 6$ matrices with one row and one column zero. 

The map described above can define $d+1$ different $\hat{\sigma}^{(r)}$ matrices for $d$ odd, and $(d+2)(d+1)/2$ different $\hat{\sigma}^{(r)}$ for $d$ even.
\item
Given one of the $l\times l$ extended matrices $\hat{\sigma}^{(r)}$, one can interpret it as a density matrix in the basis of a qubit system times a spin system with $n=2s_2+1$, where $l=2n$.  In other words, the basis is the direct product of the qubit basis $\vert m_1=\pm 1/2 \rangle$ times the $n=l/2$  basis \{ $\vert m_2=-l/2\rangle$, $\vert m_2=-l/2+1\rangle$,$\ldots$, $m_2=\vert l/2-1 \rangle$, $\vert m_2=l/2 \rangle$\}. From this assumption one can calculate the reduced density matrices for each system which correspond to a $2\times2$ and $n\times n$ matrices respectively. 
\item
If $n>2$ we repeat steps 1. and 2. for every one of the $n$-dimensional reduced density matrices obtained above. If their reduced density matrices after the implementation of the previous steps have a dimension larger than 2 we repeat the procedure, until we have only qubit states as reduced density matrices. 
\end{enumerate}

The number of total possible qubits $n_d$ defined by the procedure discussed above for a $d$-dimensional qudit system ($d\geq 4$) can be calculated as follows: The number of possible ways to extend the qudit density matrix is $d+1$ for odd $d$, and $(d+2)(d+1)/2$ for even $d$. After the partial trace operation one obtains a qubit and another density matrix of dimension $(d+1)/2$ or $(d+2)/2$, for $d$ even or odd respectively. From this argument one can define the following recursive expression
\begin{eqnarray}
n_d=f(d) (n_{g(d)}+1) \, , \quad f(d)=\Bigg\{ \begin{array}{cc} d+1 & {\rm for} \ d \ {\rm odd} \, , \\ \frac{(d+2)(d+1)}{2} & {\rm for} \ d \ {\rm even} \, ,\end{array} \nonumber \\
g(d)=\frac{1}{2}\Bigg\{ \begin{array}{cc} d+1 & {\rm for} \ d \ {\rm odd} \, , \\ d+2 & {\rm for} \ d \ {\rm even} \, ,\end{array}
\end{eqnarray}
with the initial value $n_3=8$. It is important to mention that this is the total number of possible qubits, for all the values of $d$ we will have redundant qubits or qubits which do not have information about the qutrit density matrix entries, e. g. the states
\[
\left( \begin{array}{cc}1 & 0 \\ 0 & 0 \end{array}\right) \, , \quad \left( \begin{array}{cc}0 & 0 \\ 0 & 1 \end{array}\right) \, ,
\]
which must be discarded. The elimination of this kind of cases is in general hard to do by hand but easier using a computer.

As an example of the proposed procedure we consider a $4\times 4$ density matrix
\begin{equation}
\hat{\rho}=\left( \begin{array}{cccc}
\rho_{11} & \rho_{12} & \rho_{13} & \rho_{14} \\
\rho_{21} & \rho_{22} & \rho_{23} & \rho_{24} \\
\rho_{31} & \rho_{32} & \rho_{33} & \rho_{34} \\
\rho_{41} & \rho_{42} & \rho_{43} & \rho_{44} \\
\end{array}\right) \, ,
\end{equation}
which define a set of $6 \times 6$ extended matrices with 2 rows and 2 columns zero, e.g.,
\[
\hat{\sigma}^{(1)}=\left( \begin{array}{cc}
\rho_{4\times4} & 0_{4\times2} \\
0_{2\times4} & 0_{2\times2}
\end{array}\right) \, .
\]
Assuming that this matrix is in the basis $\vert m_1=\pm 1/2, m_2=-1,0,1\rangle $ one can calculate the reduced density matrices for $s_1=1/2$ and $s_2=1$, and obtain
\begin{eqnarray*}
\hat{\sigma}(1/2)=\left( \begin{array}{cc}
\rho_{11}+\rho_{22}+\rho_{33} & \rho_{14} \\
\rho_{41} & \rho_{44}
\end{array}\right) \, , \\
\hat{\sigma}(1)=\left( \begin{array}{ccc}
\rho_{11}+\rho_{44} & \rho_{12} & \rho_{13} \\
\rho_{21} & \rho_{22} & \rho_{23}\\
\rho_{31} & \rho_{32} & \rho_{33}
\end{array}\right) \, .
\end{eqnarray*}
As stated previously, the $3 \times 3$ density matrix $\hat{\sigma}(1)$ can define up to 6 different qubit density matrices which together with $\hat{\sigma}(1/2)$ makes a total of 7 qubit density matrices that can be defined for this particular $\hat{\sigma}^{(1)}$. As we have 15 different ways to define $\hat{\sigma}^{(k)}$, with $k=1,2,\ldots, 15$, we can get up to 105 qubit density matrices whose positivity conditions can be used to calculate bounds for every entry of the original $4 \times 4$ density matrix. Nevertheless, not all of these matrices are different so only a certain amount of them can be used to define these bounds. It can be checked that for $d=4$ there exist 35 different, non-trivial qubits which can be defined, for $d=5$ there are 40, and 267 for $d=6$. It is important to notice that the number of redundant or trivial qubits grows faster than the number of non redundant, non trivial ones; as its definition is limited by the partial trace operation.

In order to construct the geometrical representation of a qudit system, we need only to take $d(d-1)/2$ qubit density matrices from all the possible. This subset of matrices $\{\hat{\rho}_{j_1}, \ldots, \hat{\rho}_{j_{d(d-1)/2}}\}$ must contain all the elements of the original density matrix in order to guarantee a proper geometrical description. For every one of the qubits $k=1,\ldots,d(d-1)/2$ we calculate the Bloch vector given by ${\cal P}_k=((\hat{\rho}_{j_k})_{12}+(\hat{\rho}_{j_k})_{21},i((\hat{\rho}_{j_k})_{12}-(\hat{\rho}_{j_k})_{21}),2(\hat{\rho}_{j_k})_{11}-1)$, and plot them inside the Bloch sphere. As in the qutrit case above, one can see that the coherence of the qudit system is the sum of the coherence of the qubits $\{\hat{\rho}_{j_1}, \ldots, \hat{\rho}_{j_{d(d-1)/2}}\}$, i.e.,
\[
C(\hat{\rho})=\sum_{k=1}^{d(d-1)/2} C(\hat{\rho}_{j_k}) \, .
\]
To illustrate how the procedure given above can be used in the state reconstruction context we consider the following example.

\section{Example: Qutrit reconstruction}
In this section we analyze the positivity conditions for the qubit states within the qutrit system and use them to obtain bounds for unknown components of the qutrit state.
The positivity conditions for the qubits can be written either in terms of the determinant of the density matrix or the trace of the squared density operator, that is
\begin{equation}
0\leq \det \hat{\rho}_j \leq 1/4, \quad 1/2 \leq {\rm Tr} (\hat{\rho}_j^2) \leq 1 \, , \quad j=1,\ldots,6 \, ,
\end{equation}
together with the conditions for the qutrit density matrix $1/3\leq {\rm Tr}(\hat{\rho}^2)\leq 1$, $1/9\leq {\rm Tr}(\hat{\rho}^3)\leq 1$ and $0 \leq \det \, \hat{\rho}\leq 1/27$ can be used to find bounds to the $3\times 3$ state components in the case where only a certain amount of them are known.

To show that, we assume a situation where all the off-diagonal components of the density matrix of Eq.~(\ref{qutrit}) are unknown, i.e.,
\begin{equation}
\hat{\rho}=\left(
\begin{array}{ccc}
\rho_{11} & x & y \\
x^* & \rho_{22} & z \\
y^* & z^* & \rho_{33}
\end{array}
\right) \, ,
\end{equation}
where $\rho_{33}=1-\rho_{11}-\rho_{22}\geq 0$, and $x$, $y$, and $z$ are unknown parameters. Taking into account the nonnegativity conditions for the qubits~(\ref{qubits1}), one can obtain the following limits for the norms of the off-diagonal terms $\vert x \vert^2$, $\vert y \vert^2$, and $\vert z \vert^2$. The conditions can be expressed as
\begin{eqnarray}
1/2\leq (1-\rho_{33})^2+\rho_{33}^2+2 \vert y \vert^2 \leq 1\, , \nonumber \\
1/2 \leq (1-\rho_{22})^2+\rho_{22}^2+2 \vert x \vert^2 \leq 1 \, , \nonumber \\
1/2\leq (1-\rho_{11})^2+\rho_{11}^2+2 \vert y \vert^2 \leq 1\, , \nonumber \\
1/2 \leq (1-\rho_{22})^2+\rho_{22}^2+2 \vert z \vert^2 \leq 1 \, , \nonumber \\
1/2\leq (1-\rho_{11})^2+\rho_{11}^2+2 \vert x \vert^2 \leq 1\, , \nonumber \\
1/2 \leq (1-\rho_{33})^2+\rho_{33}^2+2 \vert z \vert^2 \leq 1 \, , 
\end{eqnarray}
these conditions give us a rough approximation to the absolute values of the off-diagonal terms. These bounds can be expressed as
\begin{eqnarray}
\max\{0,l_1-1/4,l_2-1/4\}\leq\vert x \vert^2 \leq \min\{l_1,l_2\} \, , \nonumber \\
\max\{0,l_1-1/4,l_3-1/4\}\leq\vert y \vert^2 \leq \min\{l_1,l_3\} \, , \nonumber \\
\max\{0,l_2-1/4,l_3-1/4\}\leq\vert x \vert^2 \leq \min\{l_2,l_3\} \, , \label{appr}
\end{eqnarray}
with $l_1=\rho_{11}(1-\rho_{11})$, $l_2=\rho_{22}(1-\rho_{22})$, and $l_3=\rho_{33}(1-\rho_{33})$. These conditions may satisfy the nonnegativity conditions for the qutrit state. A better approximation to the unknown values can be obtained by means of the conditions for ${\rm Tr}(\hat{\rho}^2)$, ${\rm Tr}(\hat{\rho}^3)$, and $\det (\hat{\rho})$. As an example consider that $\rho_{11}=1/6$ and $\rho_{22}=1/3$, then from the previous expressions~(\ref{appr}) we have that $0\leq \vert x \vert^2,\vert y \vert^2 \leq 5/36$, $0 \leq \vert z \vert^2 \leq 2/9$. These conditions combined with the ones for the nonnegativity of the qutrit state lead us to different possible bounds, one of them is given by the inequalities
\begin{eqnarray*}
0\leq \vert x \vert \leq \frac{1}{3\sqrt{2}}, \quad 0 \leq \vert y \vert \leq \frac{\sqrt{1-18 \vert x \vert^2}}{2\sqrt{3}}, \\
0 \leq \vert z \vert \leq \frac{\sqrt{\left(18 \vert x \vert^2-1\right) \left(12 \vert y \vert^2-1\right)}}{\sqrt{6}}-6 \vert x \vert \vert y \vert \, ,
\end{eqnarray*}
which give us a density matrix whose purity has the limits $7/18\leq {\rm Tr}\, (\hat{\rho}^2) \leq 13/18$. As we have more information of some of the other components of the density matrix we can have better bounds.

\section{Conclusions}
A procedure to decompose a general qudit density matrix into a series of qubit systems is given. This procedure makes use of the extension of a qudit system to a higher dimension given by the closest even number greater than the original system. After that, the partial trace of the system assures the definition of positive density matrices which are reduced to qubit systems.

This qubit decomposition leads us to a graphical representation. This is done using the Bloch vectors of a quorum of $d(d-1)/2$ qubit states defined from the original $d$-dimensional qudit system. In the qutrit case the geometrical representation is given by three points in the Bloch sphere. A geometrical way to visualize the purity and the coherence of the state is given. Also, a classification of the qutrit states in terms of their purity and the region within the Bloch sphere is discussed.

We show also that the positivity conditions of the qubit decomposition, result in new inequalities which the qudit matrix components must satisfy. The use of these inequalities is discussed in the context of the state reconstruction. As an example, we analyze a system with $d=3$.

Finally, we want to point out that the decomposition presented here can be of relevance in quantum computation, as all protocols known for qubits can be applied to the ones introduced in our procedure.

Since the geometric Bloch sphere parametrization of qudit states can be mapped into a triangle geometry the qubit states are expressed in terms of measurable probability distributions~\cite{chernega17a, chernega17b, chernega17c, entropy18}. Therefore, the relations and inequalities obtained in this work for qudit density matrices (e.g., for qutrit) can be expressed as relations and inequalities in terms of tomographic probabilities.

\section*{Acknowledments}
This work was partially supported by DGAPA-UNAM (under project IN101619). 
%\end{acknowledgements}


\end{document}